\documentclass[twocolumn,showpacs,amsmath,amssymb]{revtex4}
\usepackage{graphicx}
\usepackage{amsmath}
\usepackage{longtable}
\usepackage{afterpage}
\usepackage{rotating}

\begin{document}

\title{Tune-out wavelengths for the alkaline earth atoms}

\author{Yongjun Cheng$^{1,2}$, Jun Jiang$^{2}$, and J. Mitroy$^{2}$ } 
\affiliation {$^1$Academy of Fundamental and Interdisciplinary Science, Harbin Institute of Technology, Harbin 150080, People$'$s Republic of China}
\affiliation {$^2$School of Engineering, Charles Darwin University, Darwin NT 0909, Australia} 

\date{\today}

\begin{abstract}
The lowest 3 tune-out wavelengths of the four alkaline-earth atoms,
Be, Mg, Ca and Sr are determined from tabulations of matrix elements 
produced from large first principles calculations.
The tune-out wavelengths are located near the wavelengths for
$^3P^o_1$ and $^1P^o_1$ excitations. The measurement of the tune-out 
wavelengths could be used to establish a quantitative relationship 
between the oscillator strength of the transition leading to existence 
of the tune-out wavelength and the dynamic polarizability of the 
atom at the tune-out frequency.  The longest tune-out wavelengths for 
Be, Mg, Ca, Sr, Ba and Yb are 454.9813 nm, 457.2372 nm, 657.446 nm, 
689.200 nm, 788.875 nm and 553.00 nm respectively.  

\end{abstract}

 \maketitle

\section{Introduction}

The dynamic polarizability of an atom gives a measure of the 
energy shift of the atom when it is exposed to an electromagnetic 
field \cite{miller77a,mitroy10a}. For an atom in any given state, 
one can write
\begin{equation}
\Delta E \approx - \frac{1}{2} \alpha_d(\omega) F^2,
\end{equation}
where $\alpha_d(\omega)$ is the dipole polarizability of the 
quantum state at frequency $\omega$, and $F$ is a measure of 
the strength of the AC electromagnetic field. The limiting 
value of the dynamic polarizability in the $\omega \to 0$ limit 
is the static dipole polarizability.

The dynamic polarizability will go to zero for certain frequencies of the 
applied electromagnetic field. The wavelengths at which the polarizability 
goes to zero are called the tune-out wavelengths
\cite{leblanc07a,arora11a,herold12a,holmgren12a,jiang13a}. Atoms trapped 
in an optical lattice can be released by changing the wavelength of the 
trapping laser to that of the tune-out wavelength for that atom. Very recently,
tune-out wavelengths have been measured for the rubidium \cite{herold12a} and 
the potassium atoms \cite{holmgren12a}. The advantage of a tune-out wavelength 
measurement is that it is effectively a null experiment, it measures the frequency
at which the polarizability is equal to zero. Therefore it does not
rely on a precise determination of the strength of an electric field
or the intensity of a laser field.  Accordingly, it should be possible 
to measure tune-out wavelengths to high precision and proposals to 
measure the tune-out wavelengths of some atoms with one or two valence 
electrons have been advanced \cite{cronin13a}.

The present manuscript describes calculations of the three longest tune-out 
wavelengths for Be, Mg, Ca and Sr. The tune-out wavelengths for the 
alkaline-earth atoms arise as a result of the interference between the 
dynamic polarizability coming from a weak transition and a large background 
polarizability. The tune-out wavelengths typically occur close to the 
excitation energy of the weak transitions. The atomic parameters that determine 
the values of the longest tune-out wavelengths are identified. The calculations 
utilize tables of matrix elements from earlier calculations of polarizabilities 
and dispersion coefficients \cite{mitroy10d,mitroy07e,mitroy08a,mitroy08g,mitroy10b}.
These were computed using a non-relativistic semi-empirical fixed core approach that 
has been applied to the description of many one and two electron atoms
\cite{mitroy88d,mitroy03f,mitroy09a,mitroy09b}.  In addition, the longest 
tune-out wavelengths for Ba and Yb are determined by making recourse 
to previously determined polarizabilities and oscillator strengths.

\section{Formulation}

The transition arrays for the alkaline earth atoms are essentially those
used in previous calculations of the polarizabilities and dispersion 
coefficients for these atoms 
\cite{mitroy10d,mitroy07e,mitroy08a,mitroy08g,mitroy10b}.
These were computed with a frozen core configuration interaction (CI) method. 
The Hamiltonian for the two active electrons is written
\begin{eqnarray}
H  &=&  \sum_{i=1}^2 \left(  -\frac {1}{2} \nabla^2_i
 + V_{\rm dir}({\bf r}_i) + V_{\rm exc}({\bf r}_i) +  V_{\rm p1}({\bf r}_i) \right) \nonumber \\
&+&  V_{\rm p2}({\bf r}_1,{\bf r}_2) + \frac{1}{r_{12}} \ .
\end{eqnarray}
The direct, $V_{\rm dir}$, and exchange, $V_{\rm exc}$, interactions of
the valence electrons with the Hartree-Fock (HF) core were calculated
exactly. The $\ell$-dependent polarization potential, $V_{\rm p1}$, was
semi-empirical in nature with the functional form
\begin{equation}
V_{\rm p1}({\bf r})  =  -\sum_{\ell m} \frac{\alpha_{\rm core} g_{\ell}^2(r)}{2 r^4}
      |\ell m \rangle \langle \ell m| .
\label{polar1}
\end{equation}
The coefficient, $\alpha_{\rm core}$, is the static dipole polarizability
of the core and $g_{\ell}^2(r) = 1-\exp \bigl($-$r^6/\rho_{\ell}^6 \bigr)$
is a cutoff function designed to make the polarization potential finite
at the origin. The cutoff parameters, $\rho_{\ell}$, were initially tuned to
reproduce the binding energies of the corresponding alkaline-earth positive 
ion, e.g. Mg$^+$. Some small adjustments to the $\rho_{\ell}$ were made in the 
calculations of alkaline-earth atoms to further improve agreement with the 
experimental spectrum.

A two body polarization term, $V_{p2}$ was also part of the Hamiltonian 
\cite{hameed72a,norcross76a,mitroy88d,mitroy03f}. The polarization of the 
core by one electron is influenced by the presence of the second valence 
electron. Omission of the two-body term would typically result in a $ns^2$ 
state that would be too tightly bound. The two body polarization potential 
is adopted in the present calculation with the form
\begin{equation}
V_{\rm p2}({\bf r}_i,{\bf r}_j) = -\frac{\alpha_{\rm core}} {r_i^3 r_j^3}
({\bf r}_i\cdot{\bf r}_j)g_{\rm p2}(r_i)g_{\rm p2}(r_j)\ ,
\label{polar2}
\end{equation}
where $g_{p2}$ had the same functional form as $g_{\ell}(r)$.
The cutoff parameter for $g_{\rm p2}(r)$ is usually chosen by 
averaging the different one-electron cutoff parameters.  

The use of a fixed core model reduced the calculation of the
alkaline-earths and their excited spectra to a two electron
calculation. The two electron wavefunctions were expanded in 
a large basis of two electron configurations formed from a 
single electron basis mostly consisting of Laguerre Type Orbitals. 
Typically the total number of one electron states would range 
from 150 to 200. The use of such large basis sets means that 
the error due to incompleteness of the basis is typically very small.

Details of the calculations used to represent Be, Mg, Ca and Sr have
been previously described \cite{mitroy10d,mitroy07e,mitroy08a,mitroy08g,mitroy10b}.
We refer to these semi-empirical models of atomic structure as the
configuration interaction plus core polarization (CICP) model in the subsequent
text.  

For Be, the matrix element list is exactly the same as the matrix element 
list used in Ref.~\cite{mitroy10d}.  However, the energies of the low-lying 
$2s2p \ ^{1,3}P^o$ states were set to the experimental binding energies.  In 
the case of the triplet state, the energy chosen was that of the $J = 1$ 
spin-orbit state.  Using experimental energies is important for tune-out 
wavelength calculations since the tune-out wavelength depends sensitively 
on the precise values of the excitation energies of nearby excited states.  
In the case of Mg and Ca, the reference matrix elements were those used in 
dispersion coefficient calculations \cite{mitroy07e,mitroy08a,mitroy08g}.   
The energies of the low lying Mg and Ca excited states were also set to 
experimental values for calculations of the tune-out wavelengths.   

The matrix element set for Sr incorporated experimental information.
An experimental value was used for the $5s^2$ $^1$S$^e$-$5s5p$ $^1P^o$
matrix element \cite{yasuda06a} and the energy differences for the low-lying
excitations were set to the experimental energies.  This matrix element 
set was used to calculate dispersion coefficients between two strontium atoms,  
and also between strontium and the rare gases \cite{mitroy10b}.

\begin{table}
\caption{\label{energy1} Theoretical and experimental energy levels (in Hartree)
for some of the low-lying states of alkaline-earth metals. The energies are given
relative to the energy of the core. The experimental data were taken from
the National Institute of Science and Technology (NIST) tabulation 
\cite{nistasd500} and for triplet states are the 
energies of the $J = 1$ state.}
\begin{ruledtabular}
\begin {tabular}{lcc}
  State      &   Experiment &  CICP \\
\hline
\multicolumn{3}{c}{Be}   \\
$2s^2$ $^1S^e_0$ & $-$1.0118505  & $-$1.0118967  \\
$2s2p$ $^1P^o_1$ & $-$0.8179085  & $-$0.8178898  \\
$2s3p$ $^1P^o_1$ & $-$0.7376168  & $-$0.7376426  \\
$2s2p$ $^3P^o_1$ & $-$0.9117071  & $-$0.9116666  \\
$2s3p$ $^3P^o_1$ & $-$0.7434484  & $-$0.7433848  \\
\multicolumn{3}{c}{Mg}   \\
$3s^2$ $^1S^e_0$ & $-$0.8335299  & $-$0.8335218      \\
$3s3p$ $^1P^o_1$ & $-$0.6738246  & $-$0.6737887     \\
$3s4p$ $^1P^o_1$ & $-$0.6086897  & $-$0.6086551     \\
$3s3p$ $^3P^o_1$ & $-$0.7338807  & $-$0.7336286     \\
$3s4p$ $^3P^o_1$ & $-$0.6155347  & $-$0.6156088     \\
\multicolumn{3}{c}{Ca}   \\
$4s^2$ $^1S^e_0$ & $-$0.6609319  & $-$0.6609124    \\
$4s4p$ $^1P^o_1$ & $-$0.5531641  & $-$0.5531844    \\
$4s5p$ $^1P^o_1$ & $-$0.4935704  & $-$0.4934062    \\
$4s6p$ $^1P^o_1$ & $-$0.4710284  & $-$0.4706060    \\
$4s4p$ $^3P^o_1$ & $-$0.5916298  & $-$0.5913732    \\
$4s5p$ $^3P^o_1$ & $-$0.4943762  & $-$0.4948801    \\
$3d4p$ $^3P^o_1$ & $-$0.4817070  & $-$0.4815337   \\
\multicolumn{3}{c}{Sr}   \\
$5s^2$ $^1S^e_0$ & $-$0.6146377  & $-$0.6146378   \\
$5s5p$ $^1P^o_1$ & $-$0.5157723  & $-$0.5157723   \\
$5s6p$ $^1P^o_1$ & $-$0.4592740  & $-$0.4591686   \\
$5s5p$ $^3P^o_1$ & $-$0.5485511  & $-$0.5476478   \\
$5s6p$ $^3P^o_1$ & $-$0.4603223  & $-$0.4600070   \\
$4d5p$ $^3P^o_1$ & $-$0.4446740  & $-$0.4446176   \\
\end{tabular}
\end{ruledtabular}
\end{table}

\subsection{Energies}

The energy levels of ground state and some of the lowest energy 
$^{1,3}P^{o}$ excited states for Be, Mg, Ca and Sr are listed in 
Table \ref{energy1}. The polarization potential cutoff parameters
were chosen to reproduce the energy of the might most tightly bound 
state of each symmetry. The energy of the second lowest state 
does not have to agree with the experimental energy. The reasonable
agreement with experimental energies for the second lowest states 
is an indication that the underlying model Hamiltonian is reliable.    

\subsection{Line Strengths}

Tables \ref{line1} and \ref{line2} give the line strengths for a number of
the low-lying transitions of the alkaline-earth metals comparing with available
experimental and theoretical information. The line strength can be calculated as
\begin{equation}
S_{ij} =  | \langle \psi_i; L_i J_i \parallel \  r^k
{\bf C}^{k}({\bf \hat{r}}) \parallel \psi_{j};L_j J_j \rangle|^2 \ .
\label{lines}
\end{equation}
The CICP values were computed with a modified transition operator 
\cite{hameed68a,hameed72a,mitroy88d},
e.g.
\begin{equation}
{\bf r} = {\bf r} - \left(1 - \exp(-r^6/\rho^6) \right)^{1/2} \frac{\alpha_d {\bf r}}{r^3}.
\label{dipole}
\end{equation}
The cutoff parameter used in Eq.~(\ref{dipole}) was taken as an average of the 
$s$, $p$, $d$ and $f$ cutoff parameters. The specific values are detailed 
elsewhere \cite{mitroy10d,mitroy07e,mitroy08a,mitroy08g,mitroy10b}.  

\begin{table*}
\caption{
\label{line1} 
Comparison of line strengths for the principal 
transitions of Be and Mg. The CIDF values are produced using the given 
oscillator strength and transition energies. The multi-configuration 
Hartree-Fock (MCHF) and many-body perturbation theory (MBPT) values 
are derived from the published reduced matrix elements.
Numbers in brackets represent the uncertainties in the last digits. 
The notation $a[b]$ means $a\times 10^b$.}
\begin{ruledtabular}
\begin {tabular}{lcccccc}
Final State   & $\Delta E$ (a.u.)  &  CICP   &  MCHF  & CIDF \cite{glowacki06a} &  MBPT  & Experiment\\
\hline
        \multicolumn{7}{c}{{\bf Be}}    \\
$2s2p$ $^1P^o_1$ &0.193942 &10.63   & 10.64 \cite{fischer04a} & 10.338 & 10.63 \cite{porsev02a} & 10.37(39) \cite{reistad86a}; 10.36(23) \cite{schnabel00a}\\
$2s3p$ $^1P^o_1$ &0.274199 & 0.0474  & 0.04911 \cite{fischer04a}& &  &  \\
$2s2p$ $^3P^o_1$ &0.100143 &       & 5.947[-8] \cite{fischer04a} &6.049[-8] & &\\
$2s3p$ $^3P^o_1$ &0.268402 &         & 3.182[-9] \cite{fischer04a} & & &  \\
        \multicolumn{7}{c}{{\bf Mg}}    \\
$3s3p$ $^1P^o_1$ &0.159705 &16.26  & 16.05 \cite{fischer06a}& 16.51 & 16.24 \cite{porsev02a}  & 17.56(94) \cite{larsson93a};  17.22(84) \cite{liljeby80a};  16.48(81) \cite{kwong82a}\\
$3s4p$ $^1P^o_1$ &0.224840 & 0.7062 & 0.7541 \cite{fischer06a} & & & \\
$3s3p$ $^3P^o_1$ &0.099649 &      &3.492[-5] \cite{fischer06a} &2.806[-5]  & 4.096[-5] \cite{porsev01a}  & 2.78(44)[-5] \cite{godone92a};  3.10(42)[-5] \cite{kwong82a}\\
$3s4p$ $^3P^o_1$ &0.217995 &      & 4.238[-7] \cite{MCHF09} & & &\\
\end{tabular}
\end{ruledtabular}
\end{table*}

There appears to be no experimental or theoretical data available for 
the strontium $5s^2 \ ^1S^e \to 5s6p \ ^3P^o_1$ transition \cite{sansonetti10a}.    
The line strength adopted for this transition was determined by 
estimating the mixing between the   
$5s6p \ ^1P^o_1$ and $5s6p \ ^3P^o_1$ states caused by the 
spin-orbit interaction.  The transition rates for the 
$5s6p \ ^1P^o_1 \to  5s4d \ ^1D^e_2$ and  
$5s6p \ ^3P^o_1 \to  5s4d \ ^1D^e_2$ have been measured 
\cite{werij92a,sansonetti10a}. The ratio of these transition rates 
can be used to make an estimate of the singlet:triplet mixing between 
the two $5s6p$ states with $J = 1$. Using the singlet:triplet 
mixing ratio, and the CICP line strength for the 
$5s^2 \ ^1S^e \to 5s6p \ ^1P^o_1$ transition, we estimate the 
$5s^2 \ ^1S^e \to 5s6p \ ^3P^o_1$ line strength to be 0.012.

\begin{table*}
\caption{\label{line2} Comparison of line strengths for the principal transitions of
Ca, Sr, Ba and Yb. The CIDF line strengths are produced using the given
oscillator strength and transition energies. The MCHF and CI+MBPT values
are determined from published reduced matrix elements.
Numbers in brackets represent the uncertainties in the last digits.
The notation $a[b]$ means $a\times 10^b$.}
\begin{ruledtabular}
\begin {tabular}{lcccccc}
Final State &$\Delta E$ (a.u.) &   CICP   & MCHF & CIDF \cite{glowacki03a} & MBPT  & Experiment\\
\hline
\multicolumn{7}{c}{{\bf Ca}}   \\
$4s4p$ $^1P^o_1$&0.107768  &24.37  &24.51 \cite{fischer03a} & & 24.31 \cite{porsev02a} & 24.67(90) \cite{zinner00a}; 24.9(4) \cite{kelly80a}\\
&&&&&& 24.12(1) \cite{vogt07a};   24.3(1.1) \cite{hansen83a} \\
$4s5p$ $^1P^o_1$&0.167362  &0.00666 & 0.0529 \cite{fischer03a}  & &  &\\
$4s4p$ $^3P^o_1$&0.069302  &       & 0.0011022 \cite{fischer03a}&  & 0.001156 \cite{porsev01a}& 0.00127(3) \cite{husain86a};   0.00124(7) \cite{drozdowski97a}; 0.00127(11) \cite{whitkop80a} \\
$4s5p$ $^3P^o_1$&0.166556  &       & 1.2423[-4] \cite{fischer03a} & & & \\
\multicolumn{7}{c}{{\bf Sr}}   \\
$5s5p$ $^1P^o_1$&0.098865  &28.07 & 32.18 \cite{vaeck91a} &28.8 &28.0 \cite{porsev02a} &27.54(2) \cite{yasuda06a}; 27.77(16) \cite{nagel05a}; \\
&& & & &27.12 \cite{safronova13b} &31.0(7) \cite{kelly80a};  29.2(9) \cite{parkinson76a} \\
$5s6p$ $^1P^o_1$&0.155364  & 0.0712 &0.0492 \cite{vaeck91a} & &0.0790 \cite{safronova13b} & 0.068(10) \cite{parkinson76a}  \\
$5s5p$ $^3P^o_1$ & 0.066087  & & &0.01718 & 0.0256 \cite{porsev01a} & 0.02280(54) \cite{yasuda06a}; 0.02206(51) \cite{kelly88a} \\
&& & & &0.0250 \cite{safronova13b} & 0.02418(50) \cite{husain84a};  0.0213(58) \cite{parkinson76a}  \\
$5s6p$ $^3P^o_1$&0.154315  &  &   &   &  & 0.012\footnotemark[1]   \\
\multicolumn{7}{c}{{\bf Ba}}   \\
$6s6p$ $^1P^o_1$&0.082289  &      &                       & 31.8  & 30.47 \cite{porsev02a} & 29.91(25) \cite{bizzarri90a} \\
                  &          & & &      & 29.92 \cite{dzuba06a} &  \\
$6s6p$ $^3P^o_1$ & 0.066087  & & &0.309  & 0.2746 \cite{dzuba06a} &  0.259(13) \cite{miles69a}\\
\multicolumn{7}{c}{{\bf Yb}}   \\
$6s6p$ $^1P^o_1$&0.098865  &      &                 & 16.9  & 22.85 \cite{safronova12f} & 17.30 \cite{enomoto07a} ; 17.206(17) \cite{takasu04a} \\
                &          &      &                  &     & 19.4(7.0) \cite{porsev99a} & \\
$6s6p$ $^3P^o_1$ & 0.066087 &    &   & 0.324 & 0.325 \cite{safronova12f} & 0.335 \cite{budick70a} \\
                 &         &     &    &      & 0.29(8) \cite{porsev99a} &   \\
\end{tabular}
\end{ruledtabular}
\footnotetext[1]{The experimental $5s6p$ $^1P^o_1$ line strength \cite{parkinson76a} was multiplied by 0.179 to allow for mixing with the $5s6p$ $^3P^o_1$ configuration}.
\end{table*}
\section{Polarizabilities}

\subsection{Static polarizabilities }

The polarizabilities for the ground states of Be, Mg, Ca,
and Sr are listed in Table \ref{polar}. All polarizabilities 
are computed using experimental energy differences for the 
lowest energy excited states.  The present polarizabilities  
are in good agreement with the previous high quality 
calculations.   
\begin{table}
\caption{\label{polar} Static dipole polarizabilities for the
alkaline-earth atom ground states. All values are in atomic units. 
Hybrid values were computed by replacing the line strength for 
the resonance transition with the best available experimental 
value.  }
\begin{ruledtabular}
\begin {tabular}{lcccc}
                       &   Be    &  Mg  &   Ca    &    Sr     \\
\hline
Present: CICP          &  37.73   &  71.39    &  159.4  &  197.8\footnotemark[1]     \\
Theory: RCCSD \cite{lim99a} & &       &  158.00   &  198.85     \\
Expt. \cite{molof74a}  &      &     &  169(17)   &  186(15)     \\
CI+MBPT \cite{porsev06a}  &  37.76   &  71.33 &   159.0 & 202.0  \\
CI+MBPT-SD \cite{safronova13b}  &    &     &      &  198.9       \\
Hybrid: Sum rule  \cite{porsev06a}  & &  &  157.1(1.3)\footnotemark[2]  &  197.2(2)\footnotemark[1] \\
\end{tabular}
\end{ruledtabular}
\footnotetext[1]{An experimental value \cite{yasuda06a} was used for the $5s^2$ $^1$S$^e$-$5s5p$ $^1P^o$
matrix element}.
\footnotetext[2]{An experimental value \cite{degenhardt03a} was used for the $4s^2$ $^1$S$^e$-$4s4p$ $^1P^o$
matrix element}.
\end{table}

These polarizabilities contain contributions from the core electrons.   
The electric dipole response of the core is described by a pseudo-oscillator 
strength distribution \cite{margoliash78a,kumar85a,mitroy03f}. 
Oscillator strength distributions have been constructed by using 
independent estimates of the core polarizabilities to constrain the 
sum rules \cite{mitroy03e,mitroy03f,mitroy04b,zhang07a}.  These 
take the form 
\begin{equation} 
\alpha_{\rm core} = \sum_i \frac{f_i}{\epsilon_i^2}
\end{equation} 
where $f_i$ is the pseudo-oscillator strength for a given 
core orbital and $\epsilon_i$ is the excitation energy for 
that orbital.  The sum of the pseudo-oscillator strengths 
is equal to the number of electrons in the atom. The 
pseudo-oscillator strength distribution is tabulated in Table 
\ref{core1}.  

The relative uncertainties in the polarizabilities are assessed at $0.1\%$ 
for Be, $0.5\%$ for Mg, $1.5\%$ for Ca and $1\%$ for Sr.    
\begin{table}
\caption{\label{core1} Pseudo-spectral oscillator strength distributions for the
Be$^{2+}$, Mg$^{2+}$, Ca$^{2+}$ and Sr$^{2+}$ cores.  Energies are given in a.u.. 
Refer to the text for interpretation.}
\begin{ruledtabular}
\begin {tabular}{lcccc}
 $i$ &    $\varepsilon_i$   & $f_i$  &    $\varepsilon_i$   & $f_i$    \\
\hline
  &   \multicolumn{2}{c}{Be$^{2+}$}  & \multicolumn{2}{c}{Mg$^{2+}$}  \\
 \cline{2-3} \cline{4-5} 
1 & 10.473672 & 1.0  &  50.576100 & 2.0     \\
2 &  4.813272 & 1.0  &   5.312100 & 2.0     \\
3 &           &      &   3.826606 & 6.0      \\
  &  \multicolumn{2}{c}{Ca$^{2+}$} & \multicolumn{2}{c}{Sr$^{2+}$}     \\
 \cline{2-3} \cline{4-5} 
1 & 149.495476 & 2.0 & 583.696195 &  2.0     \\
2 &  16.954485 & 2.0 &  80.400045 &  2.0     \\
3 &  13.761013 & 6.0 &  73.004921 &  6.0     \\
4 &   2.377123 & 2.0 &  13.484060 &  2.0     \\
5 &   1.472453 & 6.0 &  10.708942 &  6.0     \\
6 &            &     &   5.703458 & 10.0     \\
7 &            &     &   1.906325 &  2.0     \\
8 &            &     &   1.107643 &  6.0     \\
\end{tabular}
\end{ruledtabular}
\end{table}

\subsection{Dynamic polarizabilities and feasibility analysis }

\begin{figure}[th]
\caption{(color online) The dynamic polarizability of the neutral calcium atom in
the vicinity of the longest tune-out wavelength.}
\label{fig1}
\vspace{0.1cm}
\includegraphics[width=8.4cm,angle=0]{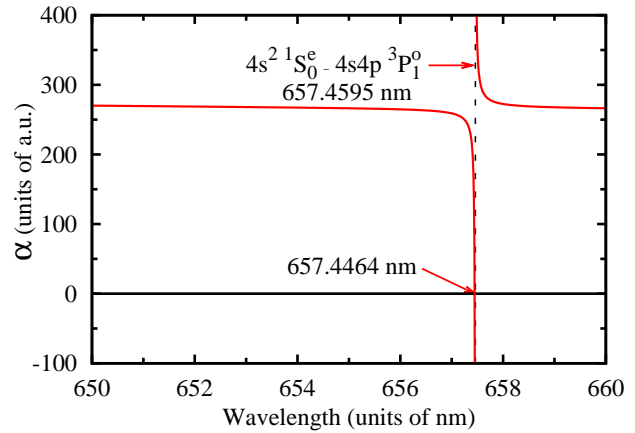}
\end{figure}

\begin{figure}[th]
\caption{(color online) The dynamic polarizability of the neutral calcium atom in
the vicinity of the second and third longest tune-out wavelengths.}
\label{fig2}
\vspace{0.1cm}
\includegraphics[width=8.4cm,angle=0]{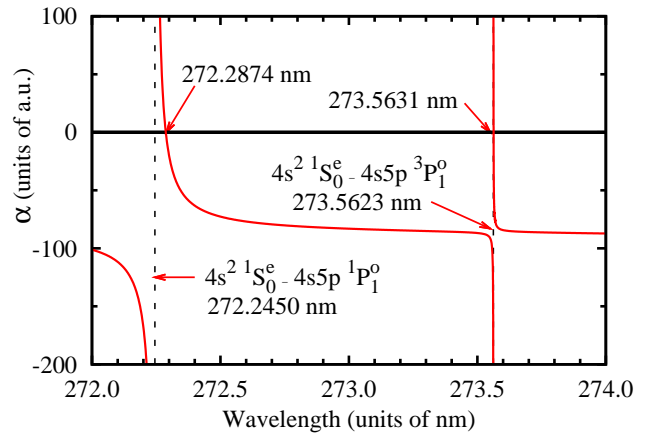}
\end{figure}

The tune-out wavelengths first require calculations of the 
dynamic polarizabilities. Some non-relativistic forbidden transition to 
the $nsnp \ ^3P^o_1$ states are included in the present calculation. 
The line strengths of these transitions are collected from MCHF calculations  
\cite{fischer03a,fischer04a,fischer06a,MCHF09} and the all-order 
MBPT calculations \cite{safronova13b}.  These line strengths are
listed in Tables \ref{tune1} and \ref{tune2}.

The dynamic polarizabilities are dominated by the $ns^2 \ ^1S^e$ 
$\to$ $nsnp \ ^1P^o_1$ resonant transition. Figures \ref{fig1} 
and \ref{fig2} show the dynamic polarizabilities of neutral 
calcium near the tune-out wavelengths and is typical of all the 
alkaline-earth atoms.  The tune-out wave 
lengths all occur close to the excitation 
energies for transitions to $^1P^o_1$ or $^3P^o_1$ states.  
The first tune-out wavelength is associated with the 
$ns^2 \ ^1S^e$ $\to$ $nsnp \ ^3P^o_1$ inter-combination transition.  
The dynamic polarizability for this transition becomes large and 
negative just after the photon energy becomes large enough to 
excite the $nsnp \ ^3P^o_1$ state.  This large negative polarizability 
will cancel with the positive polarizability from the remaining 
states at the tune-out wavelength.  The dynamic polarizability 
also has a sign change when the photon energy exceeds  the 
excitation energy for the $nsnp \ ^1P^o_1$ state.  This change 
in the polarizability is not associated with a tune-out wavelength.  
At energies larger than the  $ns^2 \ ^1S^e$ $\to$ $nsnp \ ^1P^o_1$ 
resonant transition energy the polarizability is negative.  Additional 
tune-out wavelengths occur just prior to the excitation energies of 
the higher $ns^2 \ ^1S^e$ $\to$ $^{1,3}P^o_1$ transitions.  

\begingroup
\begin{table}
\caption{\label{tune1} Breakdown of contributions to the static polarizability and 
the dynamic polarizabilities at the three longest tune-out wavelengths, $\lambda_{\rm to}$,  
for beryllium and magnesium.  The remainder term comes from all the valence 
transitions other than those specifically listed in the Table.  The uncertainty 
in the tune-out wavelength is given by $\delta \lambda_{\rm to}$.  The oscillator 
strength of the transition predominantly cancelling the polarizability 
due to the resonant transition are given in the row labelled $f$.  
The notation $a[b]$ means $a\times 10^b$.
 }
\begin{ruledtabular}
\begin {tabular}{lcccc}
     \multicolumn{5}{c}{Be}     \\
 $\lambda_{\rm to}$ (nm)  & $\infty$ & 454.9813  & 169.7578  & 166.422   \\
 $\omega_{\rm to}$ (a.u.) &  0       & 0.10014335 & 0.26840210 & 0.2737827  \\
$f$           &    & 3.970[-9]  & 5.694[-10] & 8.674[-3] \\
$\delta \lambda_{\rm to}$ (nm)  &          & 3.2[-8]  & 8.2[-10]  & 0.012   \\
\hline
$2s2p$ $^1P^o_1$    &  36.526   &    49.806  & $-$39.908    & $-$36.790   \\
$2s3p$ $^1P^o_1$    &   0.115   &     0.133  & 2.741        &    35.093     \\
$2s2p$ $^3P^o_1$    & 0.396[-6] & $-$51.070  & $-$0.640[-7] & $-$0.614[-7]    \\
$2s3p$ $^3P^o_1$    & 0.790[-8] & 0.918[-8]  & 35.518       & $-$0.195[-6] \\
Remainder           &   1.034   &     1.079  & 1.597        &     1.649   \\
$\alpha_{\rm core}$ &   0.0523   &     0.0523  & 0.0524        &     0.0524    \\
 Total              &  37.728   &       0     &  0    & 0  \\
 \hline
     \multicolumn{5}{c}{Mg}     \\
 $\lambda_{\rm to}$ (nm) & $\infty$  & 457.2372  & 209.0108   & 205.768 \\
 $\omega_{\rm to}$ (a.u.)  & 0        & 0.09964927 & 0.21799519 & 0.2214311  \\
     $ f $           &   & 2.320[-6]& 6.159[-8] & 0.1056 \\ 
 $\delta \lambda_{\rm to}$ (nm)  &          & 0.0002  & 2.57[-7]  & 0.238  \\
\hline
$3s3p$ $^1P^o_1$ &  67.878   &    111.151 & $-$78.637    & $-$73.590   \\
$3s4p$ $^1P^o_1$ &   2.094   &      2.606 &    34.922    &    69.578    \\
$3s3p$ $^3P^o_1$ & 0.234[-3] & $-$115.325 & $-$0.617[-4] & $-$0.593[-4]  \\
$3s4p$ $^3P^o_1$ & 0.130[-5] &  0.164[-5] &    40.017    & $-$0.408[-4]   \\
Remainder        &   0.939   &      1.086 &     3.215    & 3.529    \\
$\alpha_{\rm core}$ &0.481   &      0.482 &     0.483    &     0.483  \\
 Total           &  71.392   &    0 &   0.0    &  0   \\
\end{tabular}
\end{ruledtabular}
\end{table}
\endgroup


As can be seen from Figure \ref{fig1} and \ref{fig2}, the tune-out 
wavelengths for the alkaline-earth atoms arise as a result of 
the interference between the dynamic polarizability arising from a 
weak transition and a large background polarizability.  In the vicinity
of the tune-out wavelength the variation of background polarizability 
with energy will be much slower than the variation of the tune-out 
transition. The polarizability near the tune-out wavelength can be 
modelled as 
\begin{equation}  
\alpha = \alpha_0 + \frac{f}{\Delta E^2 - \omega^2} \ , 
\label{equation}  
\end{equation}  
where $\alpha_0$ is the background polarizability arising from all transitions 
except the transition near the tune-out wavelength.  The background 
polarizability is evaluated at the tune-out wavelength, $\omega_{\rm to}$.  
Setting $\alpha = 0$ gives 
\begin{equation}  
\omega_{\rm to} =  \sqrt{ \Delta E^2+ \frac{f}{\alpha_0} }  \ . 
\label{tuneout} 
\end{equation}  
When $f/\alpha_0 << \Delta E$  is obeyed, and this will generally be 
the case for the transitions discussed here, one can write  
\begin{equation}  
\omega_{\rm to} \approx \Delta E \left( 1 + \frac{f}{2\alpha_0 \Delta E^2 } \right)  \ . 
\label{tuneout2} 
\end{equation}  
Equation (\ref{tuneout2}) can be used to make estimate of the tune-out 
wavelength.  When the background polarizability is negative, the tune-out 
frequency is lower than the excitation energy of the transition triggering the 
tune-out condition.  The quotient, $f/(2 \alpha_0 \Delta E^2)$ a 
provides an estimate of the relative difference between the transition 
frequency and tune-out frequency in the vicinity of a transition.  

Equation (\ref{tuneout2}) can also be used for an uncertainty analysis.  
Setting $X_{\rm shift} = f/(2 \alpha_0 \Delta E^2)$, one has  
\begin{equation}  
\frac{\delta X_{\rm shift}}{X_{\rm shift}}  = \frac{\delta f}{f} + \frac{\delta \alpha_0}{\alpha_0} \ .  
\label{uncertainty} 
\end{equation}  
The contribution to the uncertainty in $X_{\rm shift}$ due to the uncertainty in the 
transition energy does not have to be considered at the present level
of accuracy.  

Neglecting the frequency dependence of $\alpha_0$, the variation in 
$\alpha$ with respect to variations in $\omega^2$ is 
\begin{equation}  
\frac{d\alpha}{d \omega^2}  = \frac{-f}{(\Delta E^2 - \omega^2)^2} \ . 
\end{equation}  
Writing $\omega^2 = \omega_{\rm to}^2 - \delta (\omega^2)  
= \Delta E^2 + \frac{f}{\alpha_0} - \delta (\omega^2) $ in the vicinity 
of $\omega_{\rm to}$ gives 
\begin{equation}  
\frac{d\alpha}{d \omega^2}  = \frac{-f}{\left( \frac{f}{\alpha_0} - \delta (\omega^2) \right)^2} \ . 
\end{equation}  
At $\omega = \omega_{\rm to}$, $\delta(\omega^2) = 0$, and one has 
\begin{equation}  
\frac{d\alpha}{d \omega^2}  = \frac{-\alpha_0^2}{f} \ , 
\end{equation}  
or 
\begin{equation}  
\frac{d\alpha}{d \omega}  = \frac{-2 \omega_0 \alpha_0^2}{f} \approx  \frac{-2 \Delta E \alpha_0^2}{ f}  
\ . 
\end{equation}  
The variation of the polarizability with $\omega$ is inversely proportional 
to the oscillator strength of the tune-out transition.   
Let us suppose that the condition for determination of the tune-out wavelength 
is that the polarizability be set to zero with an uncertainty of $\pm 0.1$ 
a.u. This means the photon energy should be determined with a 
frequency uncertainty of 
\begin{equation}
\Delta \omega  = \frac{ 0.1 f }{2 \Delta E \alpha_0^2 }  .
\label{tuneout3} 
\end{equation}
For Be and Mg, $\Delta \omega$ would be $7.6\times 10^{-13}$ a.u.  and 
$8.8\times 10^{-11}$ a.u. respectively.  These energy widths are very 
narrow and difficult to achieve with existing technology.  The energy 
windows for calcium and strontium would be 
$\Delta \omega = 5.2\times 10^{-10}$ a.u. and 
$\Delta \omega = 6.9\times 10^{-9}$ a.u. respectively.

\subsection{Tune-out wavelengths for Be, Mg, Ca and Sr}

Tables \ref{tune1} and \ref{tune2} list the three longest tune-out wavelengths 
for beryllium, magnesium, calcium and strontium. These are determined by 
explicit calculation of the dynamic polarizability at a series of $\omega$ 
values. The contributions of the various terms making up the dynamic 
polarizability at the tune-out wavelengths are given. The longest tune-out 
wavelength for all the atoms is dominated by two transitions, namely the 
resonance transition and the longest wavelength inter-combination transition. 
The size of the polarizability contributions from all other transitions 
relative to that coming from the resonant transitions are 2.5$\%$, 3.7$\%$,  3.6$\%$ 
and 3.7$\%$ for Be, Mg, Ca and Sr respectively at the longest tune-out 
wavelength. This dominant influence of resonant transitions means that 
a measurement of these tune-out wavelengths will result in a 
quantitative relationship between the dynamic polarizability and 
the oscillator strength for the lowest energy inter-combination 
transition.  For example, tune-out wavelengths would make it possible 
to determine the inter-combination oscillator strength given a value 
for the polarizability and/or the oscillator strength for the 
resonance transition.   

The differences between the tune-out energy and the nearest excitation 
energy can be estimated from Eq.~(\ref{tuneout}).   Values of 
$X_{\rm shift}$ for the lowest energy tune-out frequencies for Be $\to$ Sr are  
$X_{\rm shift} = 3.9\times 10^{-9}$, $1.0\times 10^{-6}$, 
$2.0\times 10^{-5}$ and $3.6\times 10^{-4}$ respectively.  
These ratios give an initial estimate of the relative precision needed in the 
wavelength to resolve the tune-out condition.  Measurement of the 
longest tune-out wavelength for beryllium requires a laser  
with a very precise wavelength. The level of precision required actually 
exceeds the precision with which the Be $2s^2 \ ^1S^e_0 \to 2s2p \ ^3P^o_1$ 
energy is given in the NIST tabulation \cite{nistasd500}. On the other hand, 
measurement of the Sr tune-out wavelength is much more feasible.       

Equation (\ref{uncertainty}) which is used to estimate the uncertainties in 
$X_{\rm shift}$, can also be used to determine the uncertainties in the tune-out 
wavelengths.   
Uncertainties in the tune-out wavelengths are given in Tables \ref{tune1}, 
 \ref{tune2} and \ref{tune3}.  

\begingroup
\begin{table}
\caption{\label{tune2} Breakdown of contributions to the static polarizability 
and dynamic polarizabilities at the three longest tune-out wavelengths for 
calcium and strontium.  The remainder term comes from all the valence 
transitions other than those specifically listed in the Table.  
The uncertainty in the tune-out wavelength is given by $\delta \lambda_{\rm to}$. 
The oscillator 
strength of the transition predominantly cancelling the polarizability 
due to the resonant transition are given in the row labelled $f$.  
The notation $a[b]$ means $a\times 10^b$. }  
\begin{ruledtabular}
\begin {tabular}{lcccc}
     \multicolumn{5}{c}{Ca}     \\
$\lambda_{\rm to}$ (nm) & $\infty$ & 657.446  & 273.563   &  272.287   \\
$\omega_{\rm to}$ (a.u.) & 0       & 0.0693035 & 0.1665552 & 0.1673360    \\
  $ f $      &     &  5.092[-5] & 1.379[-5]  & 7.431[-4] \\ 
$\delta \lambda_{\rm to}$ (nm)  &          & 0.003   & 0.005  & 0.282  \\
\hline
$4s4p$ $^1P^o_1$    & 150.734   &    257.030 & $-$108.554 & $-$106.827   \\
$4s5p$ $^1P^o_1$    &   0.027   &      0.032 &      2.760 &     86.758  \\
$4s6p$ $^1P^o_1$    &   1.097   &      1.267 &      4.757 &      4.911  \\
$4s4p$ $^3P^o_1$    &   0.011   & $-$266.414 &   $-$0.0022 &   $-$0.0022    \\
$4s5p$ $^3P^o_1$    & 0.497[-3] &  0.601[-3] &     86.039 &   $-$0.053  \\
Remainder           &   4.422   &    4.919   &     11.803 &  12.015  \\
$\alpha_{\rm core}$ &   3.160   &      3.166 &      3.197 &      3.198   \\
 Total              & 159.452   & 0 &   0 &   0   \\
 \hline
     \multicolumn{5}{c}{Sr}     \\
$\lambda_{\rm to}$ (nm) &  $\infty$ & 689.200   & 295.348    &    293.670      \\
$\omega_{\rm to}$ (a.u.) &   0       & 0.0661105 & 0.1542699    & 0.1551514  \\
  $f$           &    &1.101[-3]  &1.235[-3]   & 7.371[-3] \\
 $\delta \lambda_{\rm to}$ (nm)  &          & 0.042   & 0.011   & 0.049   \\
\hline
$5s5p$ $^1P^o_1$    & 185.788 &    336.054 & $-$129.482 & $-$127.012    \\
$5s6p$ $^1P^o_1$    &   0.305 &      0.373 & 21.763 &   111.764     \\
$4d5p$ $^1P^o_1$    &   0.734 &      0.848 & 2.777  &    2.869     \\
$5s5p$ $^3P^o_1$    &   0.252 & $-$348.600 & $-$0.057 &   $-$0.056     \\
$5s6p$ $^3P^o_1$    &   0.052 &      0.064 & 87.972  &  $-$4.772      \\
Remainder           &   4.901 &      5.430 & 11.113 &     11.292     \\
$\alpha_{\rm core}$ &   5.813 &      5.831 & 5.914 &     5.915       \\
 Total              & 197.845 &     0 & 0 & 0     \\    
\end{tabular}
\end{ruledtabular}
\end{table}
\endgroup 

Tables \ref{tune1} and \ref{tune2} also list the tune-out wavelengths 
near the $ns(n\!+\!1)\ ^{1,3}P^o_1$ excitations.  These tune-out 
wavelengths are more sensitive to polarizability contributions 
from higher transitions.  For example, about 25$\%$ of the positive 
polarizability contributions for the tune-out wavelength associated with 
the $4s5p \ ^{3}P^o_1$ excitation come from states other than the 
$4s^2 \ ^1S^e \to 4s5p \ ^{3}P^o_1$ transition.  
These tune-out wavelengths are in the ultra-violet part of the 
spectrum and would be more difficult to detect in an experiment. 

\subsection{Heavier systems, Ba and Yb}

There are two other atoms, namely, Ba and Yb with similar structures 
to those discussed earlier. The present calculational methodology cannot 
be applied to the determination of the tune-out wavelengths for these 
atoms due to relativistic effects. However, Eq.~(\ref{tuneout}) 
can be used to make an initial estimate of their longest tune-out wavelengths. 

The background polarizability, $\alpha_0$ is dominated by the 
$ns^2 \ ^1S^e_0 \to nsnp \ ^1P^o_1$ resonant transition which contributes 
more than $96\%$. The contribution to $\alpha_0$ from all other transitions, 
defined as $\alpha_{\rm rest}$, is much smaller and changes slowly when 
the frequency changes in the vicinity of the tune-out frequency. 

Assuming $\alpha_{\rm rest}$ has the same value at $\omega = 0$ and 
$\omega_{\rm to}$,
the value of $\alpha_{\rm rest}$ can be calculated as
\begin{equation}
\alpha_{\rm rest} = \alpha_d - \frac{f_{\rm resonant}}{\Delta E_{\rm resonant}^2} - \frac{f}{\Delta E^2},
\end{equation}
where $\alpha_d$ is the static polarizability of the ground states, $f_{\rm resonant}$
and $\Delta E_{\rm resonant}$ and $f$ and $\Delta E$ are the oscillator strengths and 
transition energies of the resonant transition and the transition near the tune-out wavelength respectively. 
Then the  background polarizability $\alpha_0$ can be represented as
\begin{equation}
\alpha_0 = \frac{f_{\rm resonant}}{\Delta E_{\rm resonant}^2 - \omega^2} + \alpha_{\rm rest}. 
\end{equation}
With this background polarizability, one can approximately predict the 
tune-out wavelength using Eq.~(\ref{tuneout}). 

The differences between the predicted longest tune-out wavelengths in 
this way and the values obtained 
using the exact background polarizability are only $2\times 10^{-9}$ nm, 
$3\times 10^{-6}$ nm, $3\times 10^{-5}$ nm, and $2\times 10^{-3}$ nm 
for Be $\to$ Sr which are much smaller than the uncertainties of the tune-out 
wavelengths.

\begin{table}
\caption{\label{tune3} Tune-out frequencies, $\omega_{\rm to}$, and wavelengths, 
$\lambda_{\rm to}$, for the longest tune-out wavelengths of barium and ytterbium. 
The uncertainty in the tune-out wavelength is given by $\delta \lambda_{\rm to}$.
The oscillator strengths and dipole polarizabilities adopted in the calculation 
are collected from \cite{dzuba06a,porsev06a} for barium and \cite{safronova12f} 
for ytterbium.  The contribution to the polarizability at the tune-out frequency 
due to the resonance transition is given.  The transition energies for barium and Model 2 
for ytterbium are taken from the NIST tabulation \cite{nistasd500}.  Model 2 for 
ytterbium has two low-lying strong transitions and data for both are given.  
The dynamic polarizability of $-$428.162 a.u. for Model 2 for Yb was computed 
with only the $6s^2 \ ^1S^e_0 \to 6s6p \ ^1P^o_1$ transition.  The second value 
of $\alpha_{\rm rest}$ allows for the change in the polarizability (due to the 
$6s6p \ ^3P^o_1$ transition) at the wavelength of 357.78 nm.  }
\begin{ruledtabular}
\begin {tabular}{lc}
Property  & Value  \\
\hline
     \multicolumn{2}{c}{Ba}     \\
$S_{\rm resonant} $        & 29.92   \\
$f_{\rm resonant} $        & 1.641   \\
$\alpha_d$ (a.u.)  & 273.5   \\
$\alpha_{\rm rest}$ (a.u.) & 27.92  \\
$f_{\rm resonant}/(\Delta E^2_{\rm resonant} - \omega^2_{\rm to} $ (a.u.)  & 477.772    \\
$\Delta  E$($6s^2 \ ^1S^e_0 \to 6s6p \ ^3P^o_1$) (a.u.) & 0.05757669   \\
$f$($6s^2 \ ^1S^e_0 \to 6s6p \ ^3P^o_1$) & 0.0105  \\
$\omega_{\rm to}$ (a.u.)  & 0.0577574    \\
$\lambda_{\rm to}$ (nm)  & 788.875     \\
$\delta \lambda_{\rm to}$ (nm)  &  0.295    \\
\hline
     \multicolumn{2}{c}{Yb: Model 1}     \\
$S_{\rm resonant}$ (a.u.)   &  22.85   \\
$f_{\rm resonant} $ & 1.802   \\
$\alpha_d$ (a.u.) & 141   \\
$\alpha_{\rm rest}$ (a.u.) & 9.614  \\
$f_{\rm resonant}/(\Delta E^2_{\rm resonant} - \omega^2_{\rm to}) $ (a.u.)  & 249.973   \\
$\Delta  E$($6s^2 \ ^1S^e_0 \to 6s6p \ ^3P^o_1$) (a.u.) & 0.08197762   \\
$f$($6s^2 \ ^1S^e_0 \to 6s6p \ ^3P^o_1$) & 0.0178  \\
$\omega_{\rm to}$  (a.u.)  & 0.0823938     \\
$\lambda_{\rm to}$ (nm)  & 553.00    \\ \hline 
     \multicolumn{2}{c}{Yb: Model 2}     \\
$S_{\rm resonant}$ (a.u.)    &  17.25, 5.543   \\
$f_{\rm resonant} $ & 1.314, 0.4851  \\
$\alpha_d$ (a.u.) & 141   \\
$\alpha_{\rm rest}$ (a.u.) & 9.614 , 7.726 \\
$\sum_i f_{i,\rm resonant}/(\Delta E^2_{i,\rm resonant} - \omega^2_{\rm to}) $ (a.u.)  & 256.064, $-$426.162 \\
$\Delta  E$($6s^2 \ ^1S^e_0 \to 6s6p \ ^3P^o_1$) (a.u.) & 0.08197762   \\
$\Delta  E$($6s^2 \ ^1S^e_0 \to 4f^{-1}6s^25d \ ^1P^o_1$) (a.u.) & 0.13148223  \\
$f$($6s^2 \ ^1S^e_0 \to 6s6p \ ^3P^o_1$) & 0.0178  \\
$\omega_{\rm to,1}$  (a.u.)  & 0.0823844     \\
$\lambda_{\rm to,1}$ (nm)  & 553.06   \\
$\omega_{\rm to,2}$  (a.u.)  & 0.126997     \\
$\lambda_{\rm to,2}$ (nm)  & 358.78     \\
\end{tabular}
\end{ruledtabular}
\end{table}

All the information adopted in the calculations for barium and ytterbium  are 
listed in Table \ref{tune3}. The predicted longest tune-out wavelength for 
barium was $\lambda_{\rm to} = 788.875$ nm.  
The energy window was $\Delta \omega = 3.6\times 10^{-8}$ a.u. and 
$X_{\rm shift}$ was 0.00314.  The uncertainty of the longest tune-out 
wavelengths for barium was $\delta \lambda_{\rm to} = 0.295$ nm.  The larger 
uncertainty in this tune-out wavelength was caused by the larger 
value of $X_{\rm shift}$.

Additional complications are present for ytterbium.  The values for Model 1  
reported in Table \ref{tune3} did not explicitly 
include the nearby $4f^{-1}6s^25d$ $^1P^o_1$ state in the polarizability calculation.    
This spectrum exhibits considerable mixing between the resonance $6s6p$ $^1P^o_1$ 
state and the $4f^{-1}6s^25d$ $^1P^o_1$ core excited state \cite{dzuba10a}. 
This mixing is caused by the small difference in the binding energies for the 
two states.  
This is the reason for the large difference between the CI+MBPT and experimental 
values for the resonant line strength in Table \ref{line2}.   
It has been argued that in cases such as this 
that one should use theoretical energy differences in polarizability 
calculations \cite{dzuba10a,safronova12f}.  So for our initial calculation 
of the tune-out frequency we use the CI+MBPT excitation energy for the  
resonant transition and the experimental excitation energy for the   
$6s6p$ $^3P^o_1$.  This model, which is detailed in Table \ref{tune2}  
predicts the longest tune-out wavelengths to be 
$\lambda_{\rm to} = 553.00$ nm.  The energy window, $\Delta \omega = 1.6\times 10^{-7}$ a.u.  
while $X_{\rm shift} = 0.00509$.  

Another model has been made that explicitly includes the $4f^{-1}6s^25d$ $^1P^0_1$ state 
in the polarizability calculation.   In this model the line-strength and excitation 
energy for the resonant excitation energy are set to experimental values.  The
line strength, 17.25(7),  was taken as the average of the two photoassociation line 
strengths \cite{takasu04a,enomoto07a} and its uncertainty was dervied from the 
difference of the two values and the quoted uncertainty of Ref.~\cite{takasu04a}.     
The excitation energy for the $4f^{-1}6s^25d$ $^1P^o_1$ state is set to experiment.  
The line strength for the $6s^2 \ ^1S^e_0 \to 4f^{-1}6s^25d$ $^1P^o_1$ transition was 
tuned by the requiring that the two states of Model 2 have the same polarizability as 
the resonant excitation for Model 1.  A summary of the important parameters of the 
Model 2 analysis is detailed in Table \ref{tune3}.  This model gives a tune-out 
wavelength of $\lambda_{\rm to} = 553.06$ nm.  The energy window, 
$\Delta \omega = 1.53\times 10^{-7}$ a.u. while $X_{\rm shift} = 0.00497$.   

Model 2 also allows for the existence of an additional tune-out wavelength 
located between the excitation frequencies of the $6s6p$ $^1P^0_1$ and 
$4f^{-1}6s^25d$ $^1P^o_1$ states.  This tune-out wavelength will be sensitive to 
the ratio of the respective line strengths and Model 2 predicts 
$\lambda_{\rm to} = 358.78$ nm with $X_{\rm shift} = -0.0332$.  For this 
calculation $\alpha_{\rm rest}$ was set to 7.726 a.u. by allowing for 
the frequency variation of the polarizability contribution from the 
$6s^2 \ ^1S^e_0 \to 6s6p \ ^3P^o_1$ oscillator strength.   

The complications of the structure of Yb are so severe that only indicative estimates 
of the uncertainty are possible.  For the longest tune-out frequency, we set $\delta f$, 
the uncertainty in the $6s^2 \ ^1S^e_0 \to 6s6p \ ^3P^o_1$ oscillator strength to 
1.5$\%$.  The uncertainty in the polarizability due to other transitions at the tune-out 
frequency was initially set to 0.018 \cite{safronova12f}.  To this was added 
an additional uncertainty of $0.024 = 6/250$, the difference between the Model 1 and 2 
predictions of the polarizability at the tune-out frequency.  The final uncertainty 
in the tune-out wavelength of the longest transition was 0.550 nm.    

There is little experimental information to assist in the assessment of the uncertainty of  
the tune-out wavelength near 358.78 nm.  The tune-out wavelength lies between the  
$6s^2 \ ^1S^e_0 \to 6s6p \ ^1P^o_1$ and $6s^2 \ ^1S^e_0 \to 4f^{-1}6s^25d \ ^1P^0_1$ 
transitions and its value would be largely determined by the ratio of the oscillator 
strengths to those transitions.  The uncertainty was determined by an analysis 
allowing that permitted 1.8$\%$ variations in the polarizability for the two resonant 
transitions while simultaneously admitting a $0.1/17.25 = 0.0058$ variation in the 
$6s^2 \ ^1S^e_0 \to 6s6p \ ^1P^o_1$ oscillator strength.  The uncertainty in 
$\lambda_{\rm to,2}$ was 0.23 nm.  This uncertainty should be interpreted with caution 
since the value of the tune-out wavelength is very sensitive to line strength adopted 
for the $6s^2 \ ^1S^e_0 \to 4f^{-1}6s^25d \ ^1P^0_1$ transition and this is estimated 
by an indirect method.

\section{Conclusion}

The three longest tune-out wavelengths for the alkaline-earth atoms 
from Be to Sr have been estimated from large scale configuration interaction 
calculations. The longest tune-out wavelengths for Ba and Yb have been 
estimated by using existing estimates of the polarizability and 
oscillator strengths. The longest tune-out wavelengths all occur at energies 
just above the $nsnp$ $^3P^o_1$ excitation threshold and arise due to 
negative polarizability from the $ns^2 \ ^1S^e_0 \to nsnp \ ^3P^o_1$ 
inter-combination line cancelling with the rest of the polarizability.   
The rest of the polarizability is dominated by contributions from the 
$ns^2 \ ^1S^e_0 \to nsnp \ ^1P^o_1$ resonant transition, with about 
96-97$\%$ of the polarizability arising from this transition. A 
high precision measurement of the longest tune-out wavelengths is 
effectively a measure relating the oscillator strength of the 
$ns^2 \ ^1S^e_0 \to nsnp \ ^3P^o_1$ inter-combination line to the  
polarizability of the alkaline-earth atoms. The very small oscillator 
strengths of the Be and Mg inter-combination lines might make 
a measurement of the tune-out wavelengths for these atoms difficult. The 
viability of a tune-out wavelength measurement is greater for the 
heavier calcium and strontium atoms with their stronger inter-combination 
lines. The longest wavelengths are all in the visible region.   

The second longest tune-out wavelength for all alkaline atoms occurs just before 
the excitation threshold of the $ns^2 \ ^1S^e_0 \to ns(n\!+\!1)p \ ^3P^o_1$ 
transition. Experimental detection of the second longest tune-out 
wavelength is more difficult since the oscillator strengths of the 
$ns^2 \ ^1S^e_0 \to ns(n\!+\!1)p \ ^3P^o_1$ transitions are smaller and 
the transition is in the ultra-violet.   The third longest tune-out   
wavelengths are typically triggered by the 
$ns^2 \ ^1S^e_0 \to ns(n\!+\!1)p \ ^1P^o_1$ transition.  The oscillator
strengths for the transition are about 0.1-5$\%$ the size of the resonant 
oscillator strength.   
The potential for detection of a zero in the dynamic
polarizability is larger, but once again the transition lies in the
ultraviolet region.

\begin{acknowledgments}

The work was supported by the Australian Research Council Discovery
Project DP-1092620. Dr Yongjun Cheng was supported by a grant from
the Chinese Scholarship Council.

\end{acknowledgments}


\end{document}